\documentclass[prl,twocolumn,aps,showpacs]{revtex4} 

\usepackage{epsfig}
\usepackage{graphicx}
\usepackage{color}

\begin{document}

\title{Density and spin response of a strongly-interacting Fermi gas \\ in the attractive and quasi-repulsive regime}

\author{F. Palestini, P. Pieri, and G. C. Strinati}

\affiliation{Physics Division, School of Science and Technology \\ University of Camerino, I-62032 Camerino (MC), Italy}

\begin{abstract}
Recent experimental advances in ultra-cold Fermi gases allow for exploring response functions under different dynamical conditions. 
In particular, the issue of obtaining a ``quasi-repulsive'' regime starting from a Fermi gas with an attractive inter-particle interaction while avoiding 
the formation of the two-body bound state is currently debated.
Here, we provide a calculation of the density and spin response for a wide range of temperature and coupling both in the attractive and 
quasi-repulsive regime, whereby the system is assumed to evolve non-adiabatically toward the ``upper branch'' of the Fermi gas.
A comparison is made with the available experimental data for these two quantities.
\end{abstract}

\pacs{03.75.Ss,03.75.Hh,74.40.-n,74.20.-z}
\maketitle

Ultra-cold Fermi gases represent testing systems for resolving many open issues in condensed and nuclear matter.
A key feature of these systems is that the inter-particle interaction can be varied with unprecedented flexibility through the use of Fano-Feshbach resonances from the weak- to the strong-coupling limits, which correspond to the presence of correlated and truly bound pairs, in the order. 
Recent experimental advances have also made it possible to achieve an accurate control of the temperature, in such a way that the temperature dependence of several physical quantities can be explored.

In particular, due to the diluteness condition of an ultra-cold Fermi gas, the temperature interval that can be explored ranges from about $5\%$ of the Fermi temperature $T_{F}$ up to several times $T_{F}$.
Such a wide temperature range allows alternative theoretical approaches and the corresponding results to be tested, from the temperature regime $T \gg T_{F}$ where the first few virial corrections to the free Fermi gas are relevant, down to the temperature region $T \approx T_{c}$ where the interplay of thermal and quantum fluctuations signals the presence of a superfluid phase that develops at the critical temperature $T_{c}$.

In this context, a recent experiment \cite{Zwierlein-2011} has reported values for the compressibility and spin susceptibility of a unitary Fermi gas over a wide temperature range ($0.2 \lesssim T/T_{F} \lesssim 10$), setting a benchmark for theoretical calculations that address the (static limits of the) density and spin correlation functions.

\begin{figure}[h]
\begin{center}
\includegraphics[angle=0,width=6.2cm]{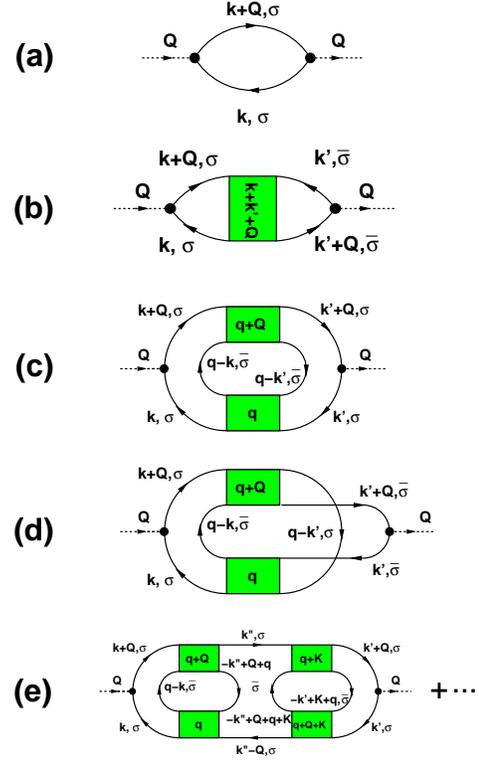}
\caption{(a) DOS, (b) MT, (c) AL , and (d) twisted AL diagrams, plus (e) series of modified AL diagrams, that are used to calculate the compressibility and spin susceptibility. The four vector Q is taken to vanish in the static limit.}
\label{fig1}
\end{center}
\end{figure}

Here, we present theoretical results for the compressibility and spin susceptibility, obtained within linear-response theory by a diagrammatic approach built 
on the t-matrix approximation and its variations.
The diagrams selected for the calculation bear on familiar contributions in condensed matter, namely, the density-of-states (DOS),
Maki-Thompson (MT), and Aslamazov-Larkin (AL) diagrams \cite{supplemental} that are depicted in Fig.÷\ref{fig1}.

In the temperature range $T_{c} \lesssim T \lesssim 5 T_{F}$ relevant to compare with the experimental data, we shall obtain the compressibility as the static limit of the density correlation function, by adding to the DOS diagram of Fig.÷\ref{fig1}(a) the MT diagram of Fig.÷\ref{fig1}(b), the AL diagrams of Figs.÷\ref{fig1}(c) and \ref{fig1}(d), plus the whole series of Fig.÷\ref{fig1}(e) which is built on these AL diagrams \cite{supplemental}.
This is because for the compressibility it is essential to take into account the residual

\noindent
interaction active above $T_{c}$ among correlated fermion pairs, in order to prevent the compressibility from diverging when approaching $T_{c}$ similarly to what occurs for point-like bosons. 
To this end, it is necessary to improve on the standard t-matrix approach, which has been successfully used in a variety of contexts but would now lead to a diverging compressibility when approaching $T_{c}$ from above.
This is achieved by including the residual interaction via the diagrammatic approach of Ref.\cite{PS-2005} which is equivalent to the Popov approach for (composite) bosons  in the BEC limit.
The good agreement we will obtain by this approach with the experimental data on the compressibility (see below), over the whole temperature range where they are available, indicates that the residual interaction among correlated fermion pairs above $T_{c}$
\cite{Stoof-2011} represents a key ingredient for the thermodynamic stability of the system.

For the spin susceptibility, only the DOS diagram of Fig.÷\ref{fig1}(a) and the MT diagram of Fig.÷\ref{fig1}(b) are relevant above (as well as below) $T_{c}$ \cite{supplemental}.           
On physical grounds, for a Fermi gas with attractive interaction one expects the spin susceptibility to be suppressed when $T_{c}$ is reached from above and to vanish eventually deep in the superfluid phase when $T \ll T_{c}$, by the argument that partners in Cooper pairs get locked in a spin singlet.                          
This feature, which can be obtained already at the mean-field level in a purely BCS approach \cite{Yosida-1958}, should be even more pronounced by the occurrence of pairing fluctuations above (as well as below) $T_{c}$ \cite{Randeria-1992}, which are associated with the occurrence of a pseudo-gap.
This behavior results also from our calculation for the attractive gas when pairing fluctuations are included, but is not consistent with  
the experimental data for the spin susceptibility of Ref.\cite{Zwierlein-2011}.

In Ref.\cite{Zwierlein-2011} the lack of suppression of the spin susceptibility close to $T_{c}$ was seen as challenging the existence of a pseudo-gap in the unitary Fermi gas \cite{Levin-2011}.
However, measurements that probe directly the single-particle excitations \cite{Jin-2008,JILA-Cam-2010} as well as a number of theoretical calculations \cite{PPSC-2002,Bulgac-2009,Ohashi-2009,Levin-2010} have supported the existence of a pseudo-gap in the unitary Fermi gas.
We shall show that the above apparent contradiction can be resolved by assuming that the measurement of the spin susceptibility of Ref.\cite{Zwierlein-2011} actually explores a non-equilibrium state associated with a \emph{quasi-repulsive} Fermi gas, and therefore cannot be directly compared with equilibrium calculations for an attractive Fermi gas, as remarked already in Ref.\cite{Taylor-2011}.           

Although the experimental data for the compressibility and spin susceptibility were reported in Ref.\cite{Zwierlein-2011} only at unitarity, we shall extend our calculations to both sides of the Fano-Feshbach resonance for couplings $(k_{F} a_{F})^{-1} > 0$ on the BEC side and $(k_{F} a_{F})^{-1} < 0$ on the BCS side of the crossover.
Here, $k_{F}$ is the Fermi wave vector related to the density by $n = k_{F}^{3}/(3 \pi^{3})$ and $a_{F}$ the scattering length.
For the attractive gas, this extension to both sides of the crossover is required to compare with the spin susceptibility data reported in Ref.\cite{Ketterle-2011-I} for the trapped gas.
For the quasi-repulsive gas, one needs to extend the calculation up to $(k_{F} a_{F})^{-1} = 10$ to recover the results of a ``dilute'' repulsive gas \cite{Galitskii-1958} with good accuracy.

\begin{figure}[t]
\begin{center}
\includegraphics[angle=0,width=9.0cm]{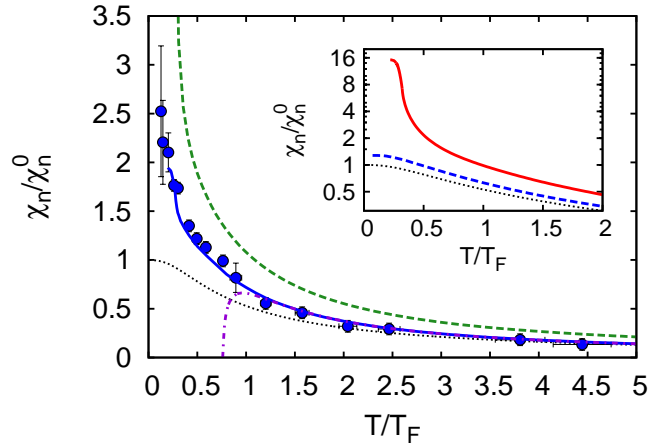}
\caption{Compressibility at unitarity vs $T/T_{F}$, normalized by the value for an ideal Fermi gas at $T=0$.
The data from Ref.\cite{Zwierlein-2011} (circles) are compared with alternative theoretical calculations. Full line: most complete calculation, including diagrams a+b+c+d+e of Fig.~1; dashed line: calculation including diagrams a+b+c+d; dashed-dotted line: virial expansion; dotted line: non-interacting Fermi gas.
The inset compares the results of our most complete calculation when $(k_{F} a_{F})^{-1} = +1.0$ (full line) and 
$(k_{F} a_{F})^{-1} = -1.0$ (dashed line) with those of a non-interacting Fermi gas (dotted line).}
\label{fig2}
\end{center}
\end{figure} 

Figure \ref{fig2} shows the temperature dependence of the compressibility at unitarity.
The experimental data from Fig.4(a) of Ref.\cite{Zwierlein-2011} (circles) are compared with the theoretical results obtained for an 
attractive Fermi gas from the static limit of the density correlation function $\chi_{n}$ in the normal phase above $T_{c}$
(in this way, only a couple of experimental data at the lowest $T$ are missed).
The calculations neglect (dashed line) or include (full line) the residual interaction among pre-formed Cooper pairs above $T_{c}$, 
and are based, respectively, on the DOS plus MT and AL diagrams and on the DOS plus MT and the \emph{whole series\/} of AL diagrams 
of Fig.\ref{fig1}.
The results obtained from the high-temperature (virial) expansion of Ref.\cite{Drummond-2009} (dashed-dotted line) and those 
for a non-interacting Fermi gas (dotted line) are also shown for comparison.
Note how the inclusion of the residual interaction among pre-formed Cooper pairs is essential to get meaningful results for the compressibility, which would otherwise diverge when $T \rightarrow T_{c}^{+}$ within the standard t-matrix approximation
\cite{PPS-unpublished}.

\begin{figure}[t]
\begin{center}
\includegraphics[angle=0,width=9.0cm]{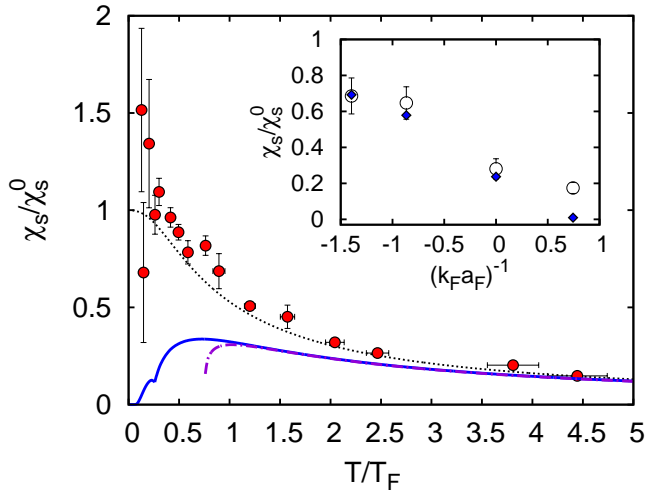}
\caption{Spin susceptibility at unitarity vs $T/T_{F}$, normalized by the value for an ideal Fermi gas at $T=0$.
The data from Ref.\cite{Zwierlein-2011} (circles) are compared with alternative theoretical calculations for an attractive Fermi gas. Full line: calculation including diagrams a+b of Fig.~1 above $T_c$ and their extension on top of the BCS contribution below $T_c$; dashed-dotted line: virial expansion; dotted line: non-interacting Fermi gas.
The inset compares the data of Ref.\cite{Ketterle-2011-I} (circles) for a trapped Fermi gas at equilibrium with our calculations (diamonds).}
\label{fig3}
\end{center}
\end{figure}  

Figure \ref{fig3} shows the corresponding temperature dependence of the spin susceptibility at unitarity.
The experimental data from Fig.4(a) of Ref.\cite{Zwierlein-2011} (circles) are compared with the theoretical results obtained for an attractive Fermi gas from the static limit of the spin correlation function $\chi_{s}$.
The results shown by the full line are obtained above $T_{c}$ by summing the contributions of the DOS [Fig.\ref{fig1}(a)] and MT [Fig.\ref{fig1}(b)] diagrams, and extended to the superfluid phase by adding the fluctuation contributions associated with these diagrams below $T_{c}$ to the BCS result \cite{Yosida-1958,Levin-2011} (see also \cite{supplemental}).
The results obtained from the high-temperature (virial) expansion of Ref.\cite{Drummond-2009} (dashed-dotted line) and those of a non-interacting Fermi gas (dotted line) are again shown for comparison \cite{footnote-cusp}.

Our results reproduce well the virial expansion for an attractive Fermi gas up to $T \approx 5  T_{F}$, and are consistently below those for a non-interacting Fermi gas.
This indicates a tendency toward pair formation in the normal state, that at lower temperature leads to a pronounced drop in the spin susceptibility which signals the presence of a ``spin gap" well before the onset of the superfluid phase \cite{Randeria-1992}.
The spin susceptibility vanishes eventually for $T \rightarrow 0$, reflecting Cooper pairing in spin singlets 
\cite{Yosida-1958,Levin-2011}.

Our results for the attractive Fermi gas, however, show marked deviations from the experimental data of Ref.\cite{Zwierlein-2011},
which lie instead above those for a non-interacting Fermi gas also at high temperature and do not perceive the expected suppression due to the singlet correlation in Cooper pairs at low temperature.
This may indicate that the specific dynamical conditions through which the spin susceptibility data were determined in Ref.\cite{Zwierlein-2011} have \emph{de facto} excluded the formation of the two-body bound state, resulting in the formation of a ``quasi-repulsive'' Fermi gas with an effective \emph{repulsive} interaction \cite{Taylor-2011}.

Before passing to describe such a quasi-repulsive Fermi gas to connect with the spin susceptibility data of Ref.\cite{Zwierlein-2011}, it may be relevant to compare our theoretical calculations for the the spin susceptibility of an attractive Fermi gas with an alternative set of data from Ref.\cite{Ketterle-2011-I}, which was taken at thermodynamic equilibrium for a trapped gas across the BCS-BEC crossover.
These data are reported in the inset of  Fig.\ref{fig3}  (circles) and are in good agreement with our theoretical calculations for the trapped system (diamonds) [the two couplings on the left (right) correspond to a temperature 
$0.13 T_{F}$ ($0.19 T_{F}$)].

A route to the description of a repulsive Fermi gas starting from an attractive one was provided recently in Ref.\cite{Ho-2011}, by the exclusion of the bound-state contribution from the density equation on the BEC side of unitarity within a Nozi\`{e}res-Schmitt-Rink (NSR) approach. 
It turns out, however, that this approach results in a wide forbidden region of the temperature-coupling phase diagram, in such a way 
that it cannot be used to obtain the spin susceptibility close to unitarity at low temperature.

Here, we propose an alternative approach to describe the quasi-repulsive Fermi gas, which focuses directly on the (particle-particle) ladder propagator 
$\Gamma_{0}$ entering the response diagrams of Fig.\ref{fig1} and eliminates the effects of the bound state on the BEC side of unitarity as follows. 
Let us consider the spectral representation 
\begin{equation}
\Gamma_{0}(\mathbf{q},i\Omega_{\nu}) = - \int_{-\infty}^{+\infty} \! \frac{d\omega}{\pi} \frac{\mathrm{Im}\Gamma_{0}^{R}(\mathbf{q},\omega)}{i\Omega_{\nu} -\omega}                  \label{Gamma_0-attractive}
\end{equation}

\noindent
for the ladder propagator of the attractive Fermi gas.
Here, $\mathbf{q}$ is the center-of-mass wave vector, $\Omega_{\nu} = 2 \pi \nu T$ ($\nu$ integer) a bosonic Matsubara frequency, and $\Gamma_{0}^{R}(\mathbf{q},\omega) = \Gamma_{0}(\mathbf{q},i\Omega_{\nu} \rightarrow \omega + i \eta)$ with $\eta = 0^{+}$.
To exclude the contribution of the two-body bound state on the BEC side of the resonance, we need to eliminate from 
$\mathrm{Im}\Gamma_{0}^{R}(\mathbf{q},\omega)$ the delta-like (polar) contribution at the given $\mathbf{q}$. 
This is done by starting the $\omega$-integration in Eq.(\ref{Gamma_0-attractive}) at the continuum threshold given by
$\omega_{c}(\mathbf{q}) = \mathbf{q}^{2}/(4m) - 2 \mu$, where $m$ and $\mu$ are the fermionic mass and chemical potential, respectively.

An analogous restriction on the frequency integration in the expression of the density is only what was required within the approach of Ref.\cite{Ho-2011} to get the thermodynamics of the quasi-repulsive Fermi gas. 
Our use of the spectral representation (\ref{Gamma_0-attractive}), however, requires us to take also into account an additional frequency-independent two-body term, which needs to be subtracted from the many-body ladder propagator in order to reproduce the correct behavior of a weakly repulsive Fermi gas when 
$(k_{F} a_{F})^{-1} \gg 1$.
This frequency-independent term can be inferred from the work of Ref.\cite{Shanenko-2007}, and yields eventually the following expression for the ladder propagator of a \emph{quasi-repulsive Fermi gas}:
\begin{equation}
\Gamma_{0}^{\mathrm{rep}}(\mathbf{p},\mathbf{q},i\Omega_{\nu}) = - \! \! \int_{\omega_{c}(\mathbf{q})}^{+\infty} \! \frac{d\omega}{\pi} \frac{\mathrm{Im}\Gamma_{0}^{R}(\mathbf{q},\omega)}{i\Omega_{\nu} -\omega}  - \frac{8 \pi / (m a_{F})}{a_{F}^{-2} + \mathbf{p}^{2}}   \!\!        \label{Gamma_0-quasi-repulsive}
\end{equation}

\noindent
where $2 \mathbf{p}$ is the incoming relative wave vector.
Note that Eqs.(\ref{Gamma_0-attractive}) and (\ref{Gamma_0-quasi-repulsive}) coincide at unitarity.
The occurrence of an energy-independent term is familiar, for instance, in the dispersion relation for the forward scattering amplitude in scattering theory \cite{LL-QM}.
We are going to use  the expression (\ref{Gamma_0-quasi-repulsive}) to calculate the thermodynamics and the response functions of this out-of-equilibrium system, provided the density equation admits solutions \cite{PPS-unpublished}.

\begin{figure}[h]
\begin{center}
\includegraphics[angle=0,width=9.0cm]{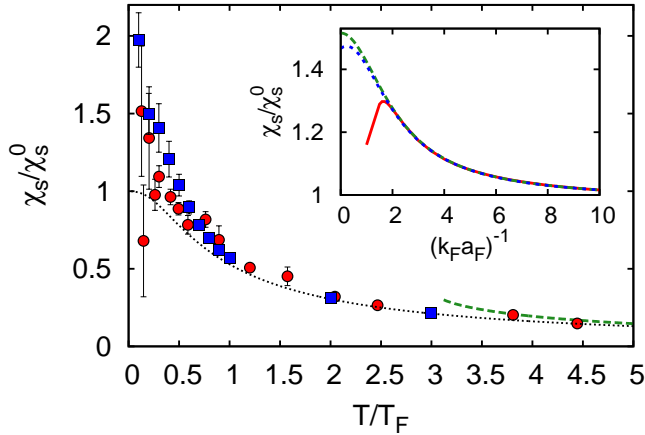}
\caption{Spin susceptibility at unitarity vs $T/T_{F}$, normalized by the value for an ideal Fermi gas at $T=0$.
The data from Ref.\cite{Zwierlein-2011} (circles) are compared with alternative theoretical calculations for a quasi-repulsive Fermi gas. Squares: our extrapolated values for a quasi-repulsive Fermi gas; dashed line: virial expansion; dotted line: non-interacting Fermi gas.
The inset shows the coupling dependence of the spin susceptibility for a given $T(=0.2 T_F)$ (full line), together with an extrapolation procedure based on two different fitting functions.}
\label{fig4}
\end{center}
\end{figure}   

In particular, we have found that the spin susceptibility at a given temperature as a function of coupling has the typical shape shown by the full line in the inset of Fig.\ref{fig4}. 
This shape coincides with that of a truly repulsive dilute Fermi gas \cite{Galitskii-1958} when $(k_{F} a_{F})^{-1} \gg 1$, and departs only slightly from it  even when the coupling gets reduced down to $(k_{F} a_{F})^{-1} \approx 2$ (this lower coupling turns out to be almost independent of temperature \cite{He-2011}). 
The spin susceptibility starts then to drop when the coupling is lowered further toward unitarity, approaching a value that correspond to an attractive Fermi gas at the given temperature (a value that can be reached only for temperatures lying outside the forbidden region in the temperature-coupling diagram).

Our assumption at this point is that, by extrapolating the shape of the spin susceptibility before it drops at $(k_{F} a_{F})^{-1} \approx 2$ in the way shown by the broken lines in the inset of Fig.\ref{fig4}  (corresponding to two different fitting functions), we should end up by reaching an excited configuration as if an avoided level crossing were present, with a dynamics determined by Landau-Zener processes \cite{Wittig-2005}.
[Similar ideas were discussed also in Refs.\cite{Demler-2011-I,Demler-2011-II} while analyzing the competing instabilities towards Stoner ferromagnetism and pairing.]
Correspondingly, we assume that the dynamics of the colliding clouds in the experiment of Ref.\cite{Zwierlein-2011} activates a number of Landau-Zener processes, such that the data there reported for the spin susceptibility at unitarity as a function of temperature can be directly compared with our extrapolated values obtained by the above procedure.

Figure \ref{fig4} compares our extrapolated values for the quasi-repulsive Fermi gas at unitarity (squares) with the data from Fig.4(a) of Ref.\cite{Zwierlein-2011} (circles) over an extended temperature range.
The error bars attached to our extrapolated values derive from the statistical uncertainties produced by the different fitting functions 
(like in the inset).
On physical grounds, these error bars may be thought of as associated with the underlying presence of a \emph{large\/} number of Landau-Zener processes mentioned in Ref.\cite{Demler-2011-II}.
Reported in the figure are also the results obtained by the high-temperature expansion of Ref.\cite{Ho-2004} where the bound-state contribution has been subtracted off (dashed line) and those of the non-interacting gas (dotted line).
The good comparison between the experimental data and our theoretical results supports our treatment of the quasi-repulsive Fermi gas as well as the assumptions about the underlying dynamical processes that result in the spin susceptibility data of Ref.\cite{Zwierlein-2011}.
It is further relevant to mention that by our approach to a quasi-repulsive Fermi gas there is no evidence for a Stoner instability toward a ferromagnetic phase, in accordance with a recent experimental finding \cite{Ketterle-2011-II}.

In conclusion, we have reported theoretical calculations for the compressibility and spin susceptibility over a wide temperature and coupling range, and compared them with recent experimental data for Fermi gases.
For the compressibility, the residual interaction among fermion pairs turned out to be an essential ingredient for comparing favorably with the experimental data for an attractive Fermi gas.
The spin susceptibility was calculated both for an attractive and a suitably defined quasi-repulsive Fermi gas, and has been favorably compared with different sets of experimental data in the two cases.
For the quasi-repulsive gas this comparison has relied on assuming a dynamics determined by Landau-Zener processes.


\clearpage


\begin{center}
{\bf Supplemental material}
\end{center}




\begin{center}
{\bf Diagrammatic formalism and linear response}
\end{center}

\noindent
The number density
\begin{equation}
\rho(\mathbf{r}) = \sum_{\alpha} \psi^{\dagger}_{\alpha}(\mathbf{r}) \, \psi_{\alpha}(\mathbf{r})
\label{number-density}
\end{equation}
\noindent
and spin density ($\hbar = 1$)
\begin{equation}
S_{z}(\mathbf{r}) = \frac{1}{2} \sum_{\alpha,\beta} \psi^{\dagger}_{\alpha}(\mathbf{r}) \, \sigma^{(z)}_{\alpha \beta}
                                \, \psi_{\beta}(\mathbf{r})
\label{spin-density}
\end{equation}
\noindent
operators, where $\sigma^{(z)}$ is a Pauli matrix,  are expressed in terms of the fermionic field operator $\psi_{\alpha}(\mathbf{r})$ at space point $\mathbf{r}$ and with spin index $\alpha = (\uparrow,\downarrow)$.
In terms of these operators, one writes the density-density
\begin{equation}
\chi_{n}(\mathbf{r} \tau,\mathbf{r'} \tau') = - \langle T_{\tau} \left[ \rho(\mathbf{r} \tau) \rho(\mathbf{r'} \tau') \right]\rangle
\label{density-density-correlation-function}
\end{equation}
\noindent
and spin-spin
\begin{equation}
\chi_{s}(\mathbf{r} \tau,\mathbf{r'} \tau') = - \langle T_{\tau} \left[ S_{z}(\mathbf{r} \tau) S_{z}(\mathbf{r'} \tau') \right] \rangle
\label{spin-spin-correlation-function}
\end{equation}
\noindent
correlation functions with imaginary time $\tau$.
Here, $T_{\tau} $ is the time-ordering operator and 
\begin{equation}
\psi^{\dagger}_{\alpha}(\mathbf{r} \tau^{+}) \, \psi_{\beta}(\mathbf{r} \tau) 
= e^{(H - \mu N)\tau} \psi^{\dagger}_{\alpha}(\mathbf{r}) \, \psi_{\beta}(\mathbf{r}) e^{-(H - \mu N)\tau}
\label{modified-Heisenberg-picture}
\end{equation}  
\noindent
is a modified Heisenberg picture with Hamiltonian $H$, number operator $N$, and chemical potential $\mu$.   
Accordingly, in Eqs.(\ref{density-density-correlation-function}) and (\ref{spin-spin-correlation-function}) the symbol $\langle \cdots \rangle$ 
corresponds to a grand-canonical average \cite{FW}.        
The correlation functions (\ref{density-density-correlation-function}) and (\ref{spin-spin-correlation-function}) represent
particular cases of the two-particle correlation function \cite{Strinati-1988}.   

From the above expressions, for a homogeneous system the (isothermal) compressibility $\chi_{n}$ and spin susceptibility $\chi_{s}$ follow as the static limits:
\begin{equation}
\chi_{n} = \left. \frac{\partial n}{\partial \mu} \right|_{T}  = \lim_{\mathbf{q} \rightarrow 0} \, \chi_{n}(\mathbf{q},\Omega_{\nu}=0)                              \label{n-static}
\end{equation}
\begin{equation}
\chi_{s} = \left. \frac{\partial M}{\partial h} \right|_{T}  = \lim_{\mathbf{q} \rightarrow 0} \, \chi_{s}(\mathbf{q},\Omega_{\nu}=0)                                  \label{s-static}
\end{equation}
\noindent
where $n$ is the density, $M$ the magnetization, $h$ a uniform magnetic field, and
\begin{eqnarray}
\chi_{n/s}(\mathbf{q},\Omega_{\nu}) & = & \int_{0}^{1/(k_{B}T)} \! d(\tau - \tau') \, e^{i \Omega_{\nu}(\tau - \tau')}  \\ 
\label{Fourier-transform-correlation-function}
& \times & \int \! d(\mathbf{r} - \mathbf{r'}) \, e^{-i \mathbf{q} \cdot (\mathbf{r} - \mathbf{r'})} \, \chi_{n/s}(\mathbf{r} \tau,\mathbf{r'} \tau') \, .\nonumber
\end{eqnarray}
\noindent
Here, $\mathbf{q}$ is a wave vector, $\Omega_{\nu} = 2 \pi \nu T$ ($\nu$ integer) a bosonic Matsubara frequency, and $k_{B}$
the Boltzmann constant.
The values of the static limits (\ref{n-static}) and  (\ref{s-static}) can conveniently be normalized in terms of the corresponding  
non-interacting values $2 N_{0}$ and $2 N_{0} \mu_{B}^{2}$, in the order, where $N_{0} = m k_{F}/(2 \pi)^{2}$
is the density of states per spin component and $\mu_{B}$ the Bohr magneton.

Standard diagrammatic methods can be used to calculate the correlation functions $\chi_{n}(\mathbf{q},\Omega_{\nu})$ and
$\chi_{s}(\mathbf{q},\Omega_{\nu})$ in Fourier space \cite{FW}, both in the normal phase above $T_{c}$ and in the superfluid phase below $T_{c}$.

\begin{center}
{\bf Choice of diagrams for \\ compressibility and spin susceptibility}
\end{center}

\begin{figure}[h]
\includegraphics[angle=0,width=8.0cm]{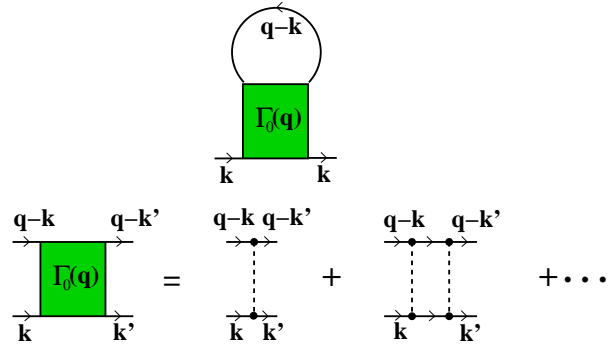}
\caption{Single-particle fermionic self-energy in the normal phase (upper panel) expressed in terms of the ladder 
              propagator $\Gamma_{0}$ between two fermions of opposite spins (lower panel).
              Full and dashed lines represent the fermionic propagator and interaction potential.}
\label{FigureS1}
\end{figure}

\noindent
Above $T_{c}$, a pairing-fluctuation approach that extends the Galitskii theory \cite{Galitskii} throughout the BCS-BEC crossover
\cite{PS-2000} identifies the relevant fermionic single-particle self-energy $\Sigma_{\mathrm{L}}$ with the diagram depicted in the upper panel of Fig.÷\ref{FigureS1}, where the particle-particle (ladder) propagator $\Gamma_{0}$ is depicted in the lower panel of the same figure. 
It sums up all the elementary scattering processes between two fermions propagating in the medium with opposite spins owing to the contact nature of the inter-particle interaction.

\begin{figure}[t]
\includegraphics[angle=0,width=4.0cm]{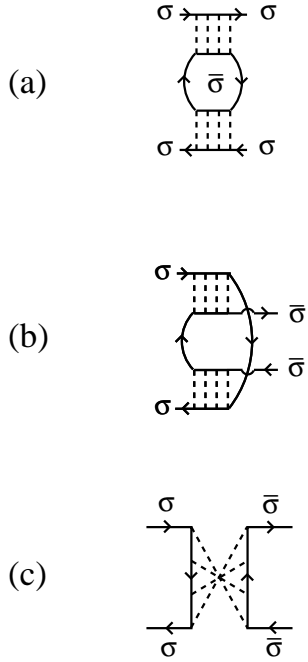}
\caption{The three types of effective two-particle interaction obtained from the pairing self-energy of Fig.÷\ref{FigureS1}, which give rise to the AL, twisted AL, and MT diagrams of Fig.1 of the main text, in the order.}
\label{FigureS2}
\end{figure}

The two-particle response that bears on this self-energy contains the effective two-particle interaction of the kinds depicted in 
Fig.÷\ref{FigureS2} (that corresponds to Fig.3 of Ref.\cite{SPL-2002}).
To the lowest order, these terms produce the Aslamazov-Larkin (AL) diagram of Fig.1(c) plus its twisted companion
of Fig.1(d) of the main text, and the Maki-Thompson (MT) diagram of Fig.1(b) of the main text.
The two AL diagrams give equal contribution to the compressibility, but cancel each other for the spin susceptibility owing to the spin structure.
On the other hand, the DOS diagram of Fig.1(a) of the main text, where the above effective two-particle interaction does not appear,
and the MT diagram of Fig.1(b) of the main text contribute to both quantities.
This justifies the choice of diagrams made in the main text.

Repeated structures based on the effective two-particle interactions of Fig.÷\ref{FigureS2} are also possible, and are specifically
required by general conservation requirements \cite{Baym-1962}.
Along these lines, for the calculation of the compressibility we have included the series of modified AL diagrams whose lowest-order contribution is depicted in Fig.1(e) of the main text.
In this case, these repeated processes are important on physical grounds because they generalize to the BCS-BEC crossover analogous processes occurring for point-like bosons in the normal phase at the Hartree-Fock level, where they introduce the effects of the mutual repulsion between bosons and thus prevent the compressibility from diverging when approaching $T_{c}$ from above.

\begin{figure}[t]
\includegraphics[angle=0,width=7.5cm]{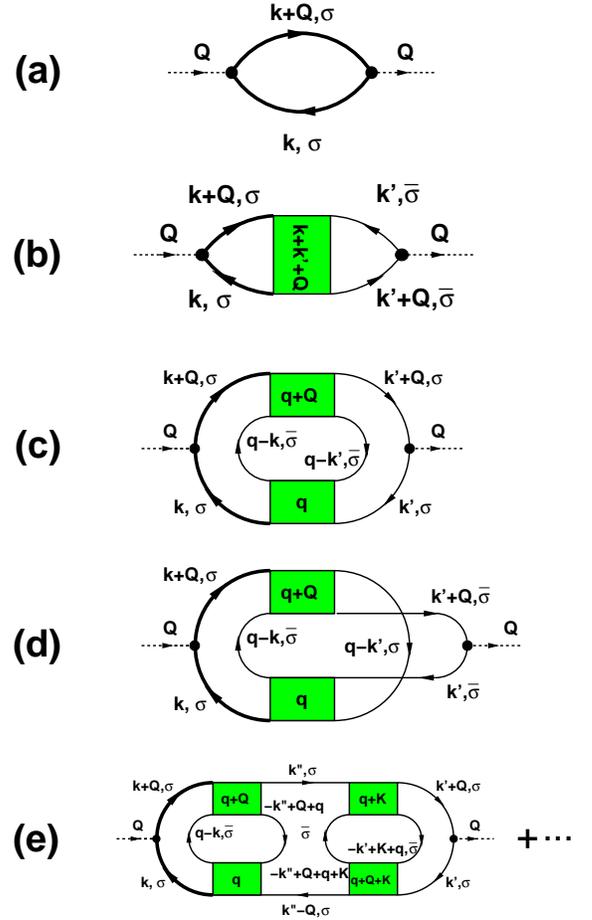}
\caption{The diagrams of Fig.1 of the main text are reproduced here with two thick lines each, which represent fermionic propagators dressed self-energy insertions.
This dressing of the fermionic lines results when calculating the compressibility as $\chi_{n} = \partial n / \partial \mu$ and
the spin susceptibility as $\chi_{s} = \partial M / \partial h$ through Ward identities.}
\label{FigureS3}
\end{figure} 

In practice, in the static limit it is possible to include this whole series of physical processes by exploiting a Ward identity that
connects single- and two-particle fermionic Green's functions \cite{LP-2006}, whereby summing the whole series of modified AL diagrams is equivalent to calculating numerically $\partial n/ \partial \mu$ in the following way.
The density is obtained from the expression
\begin{equation}
n = 2 \, k_{B} T \, \sum_{n} e^{i \omega_{n} \eta} \int \! \frac{d\mathbf{k}}{(2\pi)^{3}} \, G(\mathbf{k},\omega_{n})
\label{fermionic-density}
\end{equation}
\noindent
in terms of the fermionic single-particle Green's function
$G(\mathbf{k},\omega_{n}) = [ i \omega_{n} - \mathbf{k}^{2}/(2m) + \mu - \Sigma_{\mathrm{P}}(\mathbf{k},\omega_{n}) ]^{-1}$
where $\mathbf{k}$ is a wave vector, $\omega_{n} = \pi k_{B}T(2n+1)$ ($n$ integer) a fermionic Matsubara frequency, and 
$\eta$ a positive infinitesimal.
The fermionic self-energy $\Sigma_{\mathrm{P}}$ is formally of the type of $\Sigma_{\mathrm{L}}$ of Fig.÷\ref{FigureS1}, 
but with a dressed ladder propagator $\Gamma$ that replaces $\Gamma_{0}$ in order to include interaction processes between 
composite bosons as described by the generalized Popov theory of Ref.\cite{PP-2005}.

A comment is in order at this point about the degree of self-consistency that results when the diagrammatic structure for $\chi_{n}$
is generated in the above way by taking the derivative of the expression (\ref{fermionic-density}) for $n$ with respect to $\mu$.
One formally obtains the same diagrams of Fig.1 of the main text that contribute to $\chi_{n}$ (namely, diagrams (a), (c), (d) and the whole series (e)), but now with two fermionic propagators (identified by the thick lines in Fig.÷\ref{FigureS3}) which are dressed by the self-energy $\Sigma_{\mathrm{P}}$.

Below $T_{c}$, the calculation of the spin-spin correlation function can initially be done at the level of the BCS (mean-field) approximation, whereby the bare bubble of Fig.1(a) of the main text is replaced by the sum of two bubbles calculated, respectively, with two normal ($G_{11}$) and two anomalous ($G_{12}$) single-particle Green's functions \cite{Mahan-2000}.
In particular, in the static limit one obtains for the spin susceptibility at the BCS level \cite{Yosida-1958S}:
\begin{equation}
\frac{\chi_{s}^{(BCS)}(T)}{2 N_{0} \mu_{B}^{2}} = - \frac{1}{N_{0}} \, \int \! \frac{d\mathbf{k}}{(2\pi)^{3}} \, 
\frac{\partial f_{F}(E_{\mathbf{k}})}{\partial E_{\mathbf{k}}}
\label{Yosida-result}
\end{equation}
\noindent
where $E_{\mathbf{k}} = [ (\mathbf{k}^{2}/(2m) - \mu)^{2} + |\Delta|^{2} ]^{1/2}$ is the BCS dispersion with gap $\Delta$ and
$f_{F}(E) = (e^{E/(k_{B} T)} + 1)^{-1}$ the Fermi function.
Note that this quantity vanishes in the zero-temperature limit, reflecting the singlet structure of the Cooper pairs.
The mechanism for this to occur is a cancellation of the contributions of the normal and anomalous BCS bubbles.

Pairing fluctuations beyond mean field can be included below $T_{c}$ following the approach of Ref.\cite{PPS-2004}.
In particular, in the above two BCS (bubble) diagrams for the spin-spin correlation function the normal single-particle Green's functions $G_{11}$ are affected by pairing fluctuations while the anomalous ones $G_{12}$ remain at the BCS level. 
In addition, the MT diagram of Fig.1(b) of the main text is introduced where now all single-particle lines are $G_{11}$. 
In this way one recovers the vanishing of the spin susceptibility at zero temperature for any coupling $(k_{F} a_{F})^{-1}$.

Above $T_{c}$, on the other hand, for the spin-spin correlation function the DOS diagram replaces the normal BCS 
bubble while the MT diagram plays the role of the anomalous BCS bubble. 
In this case, a complete cancellation between the MT diagram and the fluctuation contributions to the DOS diagram occurs in the strong-coupling (BEC) limit for temperatures well below the pair-breaking temperature of the composite bosons.
This is expected on physical grounds, since a non-vanishing contribution to the spin response for spin-less composite bosons 
should result only when the temperature is comparable with their binding energy and the composite bosons break apart
\cite{SPL-2002}.

A comment on the degree of self-consistency for the diagrammatic structure is relevant also for $\chi_{s}$.
At the BCS level, no difference is introduced when calculating $\chi_{s}$ via $\partial M / \partial h$ with respect to the diagrammatic calculation resulting in the expression (\ref{Yosida-result}).
When fluctuations are introduced, while both fermionic propagators of the DOS diagram of Fig.1(a) of the main text are affected by pairing fluctuations, only the pair of fermionic propagators on the left side of the MT diagram of Fig.1(b) of the main text are dressed 
by pairing fluctuations through the self-energy $\Sigma_{\mathrm{L}}$ depicted in Fig.÷\ref{FigureS1}, both above and below $T_{c}$.
This corresponds to the occurrence of thick lines in panels (a) and (b) of Fig.÷\ref{FigureS3}. 
We have, however, verified numerically that the dressing \ref{FigureS3}(b) of the MT diagram does not affect our main physical results, namely, that $\chi_{s}$ is strongly suppressed for $T \ll T_{c}$ and vanishes at $T=0$, while it slowly decreases for increasing temperature when $T \gtrsim T_{F}$ where it coincides with the results of the high-temperature (virial) expansion.




\begin{thebibliography}{99}
\bibitem{footnote-cusp} The minor cusp present in the full curve of Fig.\ref{fig3} about \newpage $T_{c}$ originates from a spurious first-order transition between the normal and superfluid phases. Its presence, however, is irrelevant to the physical arguments raised in discussing the features of this curve 
for $T \ll T_{c}$ and $T \gtrsim T_{F}$.
\end{thebibliography}
\end{document}